\documentclass[pdftex,twocolumn,epjc3]{svjour3}          % twocolumn

\RequirePackage[T1]{fontenc}

\smartqed  % flush right qed marks, e.g. at end of proof

\RequirePackage{graphicx}
\RequirePackage{mathptmx}      % use Times fonts if available on your TeX system
\usepackage{amssymb}
\usepackage{amsmath}
\usepackage{changes}

\RequirePackage{flushend}
\RequirePackage[numbers,sort&compress]{natbib}
\journalname{Eur. Phys. J. C}

\begin{document}

\title{Gravitational waves from pulsars with measured braking index}


\author{Jos\'e C. N. de Araujo\thanksref{e1}
        \and
        Jaziel G. Coelho\thanksref{e2}
        \and Cesar A. Costa\thanksref{e3}
}

%\thankstext[$\star$]{t1}{Thanks to the title}
\thankstext{e1}{e-mail:jcarlos.dearaujo@inpe.br}
\thankstext{e2}{e-mail:jaziel.coelho@inpe.br}
\thankstext{e3}{e-mail:cesar.costa@inpe.br}
\institute{Divis\~ao de Astrof\'isica, Instituto Nacional de Pesquisas Espaciais, Avenida dos Astronautas 1758, S\~ao Jos\'e dos Campos, 12227--010 SP, Brazil\label{addr1}}

\date{Received: date / Accepted: date}

\maketitle

\begin{abstract}
We study the putative emission of gravitational waves (GWs) in particular for pulsars with measured braking index. We show that the appropriate combination of both GW emission and magnetic dipole brakes can naturally explain the measured braking index, when the surface magnetic field and the angle between the magnetic dipole and rotation axes are time dependent. Then we discuss the detectability of these very  pulsars by aLIGO and the Einstein Telescope. We call attention to the realistic possibility that aLIGO can detect the GWs generated by at least some of these pulsars, such as Vela, for example.
\end{abstract}

\section{Introduction}
\label{int}
Recently, gravitational waves (GWs) have finally been  detected~\citep{2016PhRvL.116f1102A}. The signal was identified as coming from the final fraction of a second of the coalescence of two black holes (BHs), which resulted in a spinning remnant black hole. Such event had been predicted \citep[see e.g.,][]{2010ApJ...715L.138B} but never been observed before.

As it is well known, pulsars (spinning neutron stars) are promising candidates for producing GW signals which would be detectable by aLIGO (Advanced LIGO) and AdV Virgo (Advanced Virgo). These sources might generate continuous GWs whether they are not perfectly symmetric around their rotation axes. 

The so called braking index (n), which is a quantity closely related to the pulsar spindown, can provide information about the pulsars' energy loss mechanisms. Such mechanisms can include GW emission, among others. 

Until very recently, only eight of the $\sim 2400$ known pulsars have braking indices measured accurately. All these braking indices are remarkably smaller than the canonical value $(n = 3)$, which is expected for pure magneto-dipole radiation model
\citep[see e.g.,][]{1993MNRAS.265.1003L,1996Natur.381..497L,2007ApSS.308..317L,2011MNRAS.411.1917W,2011ApJ...741L..13E,2012MNRAS.424.2213R, 2015ApJ...810...67A}. 

Several interpretations for the observed braking indices have been put forward, like the ones that propose either accretion of fall-back material via a circumstellar disk \citep{2016MNRAS.455L..87C}, relativistic particle winds \citep{2001ApJ...561L..85X,2003A&A...409..641W}, or modified canonical models to explain the observed braking index ranges \citep[see e.g.,][and references therein for further models]{1997ApJ...488..409A,2012ApJ...755...54M,2016ApJ...823...34E}. Alternatively, it has been proposed that the so-called quantum vacuum friction (QVF) effect in pulsars can explain several aspects of their phenomenology~\citep{2016ApJ...823...97C}. 
However, so far no developed model has yet explained satisfactory all measured braking indices, nor any of the existing models has been totally ruled out by current data. Therefore, the energy loss mechanisms for pulsars are still under continuous debate.

Recently, Archibald et al.~\cite{2016ApJ...819L..16A} showed that PSR J1640-4631 is the first pulsar having a braking index greater than three, namely $n=3.15\pm 0.03$. PSR J1640-4631 has a spin period of $P=206$~ms and a spindown rate of $\dot P=9.758(44)\times 10^{-13}$~s/s, yielding a spindown power $\dot E_{\rm rot}=4.4\times 10^{36}$~erg/s, and an inferred dipole magnetic field $B_0=1.4\times10^{13}$~G. This source was discovered by using X-ray timing observations with NuStar and a measured distance of $12$ kpc~\citep[see][]{2014ApJ...788..155G}.

The braking index of PSR J1640-4631 reignites the question about energy loss mechanisms in pulsars. With the exception of this pulsar, all other eight, as previously mentioned, have braking indices $n < 3$  (see Table~\ref{ta1}), which may suggest that other spindown torques act along with the energy loss via dipole radiation.
Recently, we showed that such a braking index can be accounted for if the spindown model combines magnetic dipole and GW brakes~\citep[see][]{2016arXiv160305975D}. Therefore, each of these mechanisms alone can not account for the braking index found for PSR J1640-4631.

Since pulsars can also spindown through gravitational emission associated to asymmetric deformations~\citep[see e.g.,][]{1969ApJ...157.1395O,1969ApJ...158L..71F}, it is 
appropriate to take into account this mechanism in a model which aims to explain the braking indices which have been measured.
Thus, our interest in this paper is to explore both gravitational and electromagnetic contributions in the context of pulsars with braking indices measured accurately. Based on the above reasoning, the aim of this paper is to extend the analysis of~\cite{2016arXiv160305975D} for all pulsars of Table~\ref{ta1}. In the next section we revisit the fundamental energy loss mechanisms for pulsars. We also derive its associated energy loss focusing mainly on the energy balance, when both gravitational and classic dipole radiation are responsible for the spindown. Also, we elaborate upon the evolution of other pulsars' characteristic parameters, such as the surface magnetic field $B_0$ and magnetic dipole direction $\phi$. In Section~\ref{sec3} we briefly discuss the detectability of these pulsars by  aLIGO and the planned Einstein Telescope (ET) in its more recent design (ET-D).
Finally, in Section~\ref{sec4}, we summarize the main conclusions and remarks. We work here with Gaussian units.

\section{Modeling pulsars' braking indices}\label{sec1}
\begin{table*}
\caption{{\bf{The periods ($P$) and their first derivatives ($\dot P$) for pulsars with known braking indices ($n$) (see also ATNF catalog~\citep{ATNF,2005AJ....129.1993M}).}}}
\label{ta1}
\begin{tabular*}{\textwidth}{@{\extracolsep{\fill}}lrrrrl@{}}
\hline
Pulsar & $P$~(s) &$\dot{P}~(10^{-13}$~s/s) &n & Ref.   \\ 
\hline
PSR J1734-3333      &1.17  &22.8 &$0.9\pm0.2$       &~\cite{2011ApJ...741L..13E}  \\
PSR B0833-45 (Vela) &0.089 &1.25 & $1.4\pm0.2$      &~\cite{1996Natur.381..497L}  \\
PSR J1833-1034      &0.062 &2.02 &$1.8569\pm0.0006$ &~\cite{2012MNRAS.424.2213R}  \\
PSR J0540-6919      &0.050 &4.79 &$2.140\pm0.009$   &~\cite{2007ApSS.308..317L}   \\ %(B0540-69)
PSR J1846-0258      &0.324 &71   &$2.19\pm0.03$     &~\cite{2015ApJ...810...67A}  \\
PSR B0531+21 (Crab) &0.033 &4.21 &$2.51\pm0.01$     &~\cite{1993MNRAS.265.1003L}  \\
PSR J1119-6127      &0.408 &40.2 &$2.684\pm0.002$   &~\cite{2011MNRAS.411.1917W}  \\
PSR J1513-5908      &0.151 &15.3 &$2.839\pm0.001$   &~\cite{2007ApSS.308..317L}   \\ % (B1509-58)
PSR J1640-4631      &0.207 &9.72 &$3.15\pm0.03$     &~\cite{2016ApJ...819L..16A}  \\
\hline
\end{tabular*}
\end{table*}

If the pulsar magnetic dipole moment is misaligned with respect to its spin axis by an angle $\phi$, the energy per second emitted by the rotating magnetic dipole is \citep[see e.g.,][]{1975ctf..book.....L, 2001thas.book.....P},
\begin{equation}
\dot{E}_{\rm d}= -\frac{8\pi^4}{3}\frac{B_0^2 R^6\sin^2\phi}{c^3}f_{\rm rot}^4,  \label{Ed}
\end{equation}
where $B_0$ is taken as the surface magnetic field (coming from a magnetic dipole) of a star of radius $R$
and rotational frequency $f_{rot}$, and $c$ is the speed of light.

A body with moment of inertia $I$ and equatorial ellipticity $\epsilon$ emits GWs, and the luminosity associated with this emission reads~\citep{1983bhwd.book.....S}
\begin{equation}
\dot{E}_{\rm GW} = -\frac{2048\pi^6}{5}\frac{G}{c^5}I^2\epsilon^2 f_{\rm rot}^6. \label{EGW}
\end{equation}
The upper limit for the GW strain from isolated pulsars, known as the spindown limit, can be calculated by assuming that the observed loss of rotational energy ($\dot E_{\rm rot} = 4  \pi^2 I f_{\rm rot} \dot f_{\rm rot}$) is all due to gravitational radiation $\dot{E}_{\rm GW}$.

Instead, we consider in this paper that the total energy of the star is provided by its rotational energy, $E_{\rm rot} = 2  \pi^2If_{\rm rot}^2$, and any change on it is attributed to both $\dot{E}_{\rm d}$ and $\dot{E}_{\rm GW}$. Therefore, the energy balance reads

\begin{equation}
\dot{E}_{\rm rot}\equiv \dot{E}_{\rm GW} +\dot{E}_{\rm d} \label{Erotdef},
\end{equation}
consequently, it follows immediately that 

\begin{equation}
\dot{f}_{\rm rot} = - \frac{512\pi^4}{5} \frac{G}{c^5}I\epsilon^2f^5_{\rm rot} - \frac{2\pi^2}{3}\frac{B_0^2R^6\sin^2\phi}{Ic^3}f^3_{\rm rot}. \label{domeg}
\end{equation}

This equation can be interpreted as follows: the term on the left side stands for the resulting deceleration (spindown) due to magnetic dipole and GWs brakes, the first and the second terms on the right side denote the independent contributions of these decelerating processes, respectively. Equation \ref{domeg} can be rewritten in the following form

\begin{equation}
\dot{f}_{\rm rot} = \dot{f}_{\rm GW} + \dot{f}_{\rm d}. 
\end{equation}

Now, we can define the fraction of deceleration related to GW emission, namely
\begin{equation}
\eta \equiv \frac{\dot{f}_{\rm GW}}{\dot{f}_{\rm rot}}, \label{eta0}
\end{equation}
which, by replacing the appropriate quantities, reads

\begin{equation}
\eta = \frac{1}{1+\frac{5}{768}\frac{c^2B_0^2R^6\sin^2\phi}{G\pi^2 I^2\epsilon^2f_{\rm rot}^2}}. \label{eta}
\end{equation}

From equation \ref{eta0}, it follows immediately that
\begin{equation}
\dot{E}_{\rm GW} = \eta \dot{E}_{\rm rot},
\end{equation}
thus $\eta$ can be interpreted as the efficiency of GWs generation.  
Now, we show that equation \ref{eta} is closely related to the equation for the braking index $n$. Recall that $n$ is given by
\begin{equation}
n = \frac{f_{\rm rot}\,{\ddot f}_{\rm rot}}{\dot{f}^2_{\rm rot}}\label{n}.
\end{equation}

Before proceeding, it is also worth recalling that a pure magnetic brake, in which a dipole magnetic configuration is adopted, leads to $n = 3$, whereas a pure GW brake leads to $ n = 5$.
From the observational point of view, the literature shows that almost all pulsars with measured braking index have $ n < 3$ (see Table \ref{ta1}). As discussed earlier, there is one exception: PSR J1640-4631 has a braking index of $ n \simeq 3.15$. Therefore, neither a pure GW brake nor a pure magnetic dipole brake are supported by the observations.
We have recently shown~\citep[see][]{2016arXiv160305975D} that the PSR J1640-4631 braking index can be accounted for a combination of GWs and magnetic dipole brake.  Alternatively, it has been proposed that magnetic quadrupolar radiation could explain the braking index $n>3$~
\citep[see e.g.,][and references therein]{2015ApJ...810...67A, 2015MNRAS.450..714P}.
On the other hand, for all other pulsars with known braking indices (see Table \ref{ta1}), additional mechanisms invoking an external torque similar to that from stellar wind~\citep{2015MNRAS.450.1990K,2016MNRAS.457.3922O} must be considered to explain the fact that $n < 3$. In this work we explore a possible way to explain brake indices $ n < 3$ considering that the magnetic field and/or the angle between the rotation and magnetic axes are time-dependent.

It is believed that magnetic fields should decay in pulsars [usually due to Ohmic decay, Hall drift and ambipolar diffusion \citep{1988MNRAS.233..875J,1992ApJ...395..250G}] on timescales of $(10^{6}-10^{7})$ yr \citep[see e.g.,][and references therein]{1992ApJ...395..250G, 2015MNRAS.453..671G}. Nevertheless, there are also suggestions that the timescales for $B_0$ could actually be smaller, of the order of $10^5$ yr \citep{2014MNRAS.444.1066I,2015AN....336..831I}.

\begin{figure}
\includegraphics[width=\linewidth,clip]{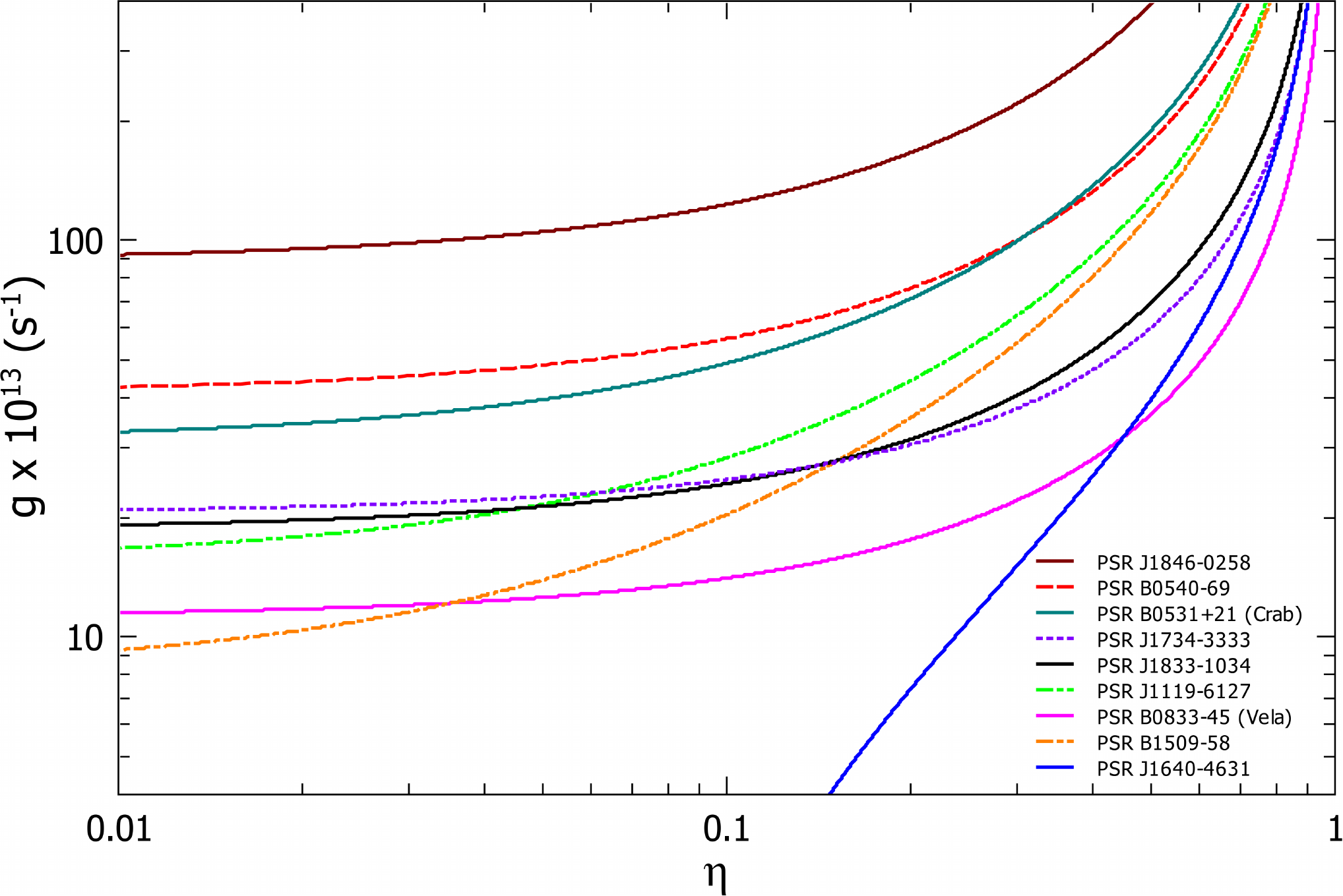}
\caption{The term in brackets ($g$) as a function of $\eta$.}\label{fig1}
\end{figure}

To proceed, by substituting equation \ref{domeg} and its first derivative in equation \ref{n}, the braking index reads
\\
\begin{equation}
n=n_0+ \frac{f_{\rm rot}}{{\dot f}_{\rm rot}}\left(5-n_0\right)\left[\frac{\dot B_0}{B_0}+\dot{\phi}\cot{\phi}  \right], \label{nBphi}
\end{equation}
with 
\begin{equation}
n_0 = 3 + \frac{2}{1+\frac{5}{768}\frac{c^2B_0^2R^6\sin^2\phi}{G\pi^2 I^2\epsilon^2f_{\rm rot}^2}},
\end{equation}
where we consider that $\dot{B_0}$ and $\dot{\phi}$ are not null. Also, substituting equation \ref{eta} into equation \ref{nBphi} we finally obtain an equation that relates the braking index to the efficiency of GWs generation, namely
\begin{equation}
n=3+2\eta-2\frac{P}{\dot P}\left(1-\eta\right)\left[\frac{\dot B_0}{B_0}+\dot{\phi}\cot{\phi}  \right],
\end{equation}
\citep[see][for a similar analysis]{2000A&A...354..163P} conveniently written in terms of the rotational period ($P = 1/f_{\rm rot}$) and its first derivative ($\dot{P}$), in order to be directly applied to the data of Table \ref{ta1}. Notice that for a given $\eta$, the above equation shows that in principle it is possible to obtain $ n < 3$ if the appropriate combination of $\dot{B_0}$ and $\dot{\phi}$ turns the term in brackets positive.
In order to proceed, it is interesting to calculate the term in brackets as a function of $\eta$ for the pulsars of Table \ref{ta1}. For the sake of simplicity, the term in brackets is rewritten as follows
\begin{equation}
g = g(B_0,\dot{B}_0,\phi,\dot{\phi})\equiv \left[\frac{\dot B_0}{B_0}+\dot{\phi}\cot{\phi}  \right]. \label{gb}
\end{equation}
Thus, the term in brackets as a function of $\eta$ for a given pulsar reads
\begin{equation}
g = -\frac{(n-3-2\eta)}{2(1-\eta)}\frac{\dot{P}}{P}.
\end{equation}
In Figure \ref{fig1} we present the term in brackets ($g$) as a function of $\eta$. This figure shows that it is in principle possible, as already mentioned, to find a combination of $\dot{B_0}$ and $\dot{\phi}$ in order to have $ n < 3$ and GWs be generated. 
Thus, bearing in mind that magnetic fields for pulsars are within $(10^{12}-10^{13})$ G, let us assume $\dot{B}_0<0$ and $|\dot{B}_0|$ of the order of $(10^{-2}-10^{-1})$~G/s. Since the Crab pulsar has an observationally inferred $\dot\phi\simeq 3\times 10^{-12}$ rad/s~\citep{2013Sci...342..598L, 2015MNRAS.446..857L,2015MNRAS.454.3674Y}, let us consider the implications of these parameters. For instance, consider the representative angle $\phi = \pi/4$ , and $\dot{B}_0= -0.05$ G/s, from that we obtain $g\simeq 3\times 10^{-12}$~$\rm s^{-1}$. Figure \ref{fig1} shows that this value for $g$ implies that $\eta < 0.1$. Therefore, our model can provide a consistent picture. In addition, our results show that for reasonable values of $g$, $\eta$ can not be arbitrarily large, or vice-versa.

Notice that PSR J1640-4631 can also have its braking index $n = 3.15$ consistently explained. In our previous paper~\citep[see][]{2016arXiv160305975D} $\eta = 0.075$, in the present model we can have $0 \leq \eta \leq 1$, depending on the values of $\dot{B_0}$ and $\dot{\phi}$.

Also, it is worth mentioning that an appropriate negative value of $g$ can account for braking indices $n >$3. For example, scenarios in which the pulsar's magnetic field decays implies $g < 0$.

Finally, it is worth noting that \cite{2000A&A...354..163P} studied in particular four of the nine pulsars in Table \ref{ta1} in this present paper. Moreover, our parametrization and interest are different, since we here relate $\eta$ and $g$, which are not used and nor explicitly defined, in the referred paper, to explain the braking indices in Table \ref{ta1}.
\section{Relating $\eta$ and $\epsilon$ with the amplitude of GWs}\label{sec3}
Now, it is appropriate to consider how the GW amplitudes for pulsars with $n < 5$ can be calculated. Recall that one usually finds in the literature the following equation
\begin{equation}
h^2 = \frac{5}{2}\frac{G}{c^3}\frac{I}{r^2}\frac{\mid\dot{f}_{\rm rot}\mid}{f_{\rm rot}},
\end{equation}
\citep[see, e.g.,][]{2014ApJ...785..119A} where one is considering that the whole contribution to $\dot{f}_{\rm rot}$ comes from GW emission, which means that one is implicitly assuming that $ n = 5$. This equation must be modified to take into account that $n < 5$. 

\begin{figure}
\includegraphics[width=\linewidth,clip]{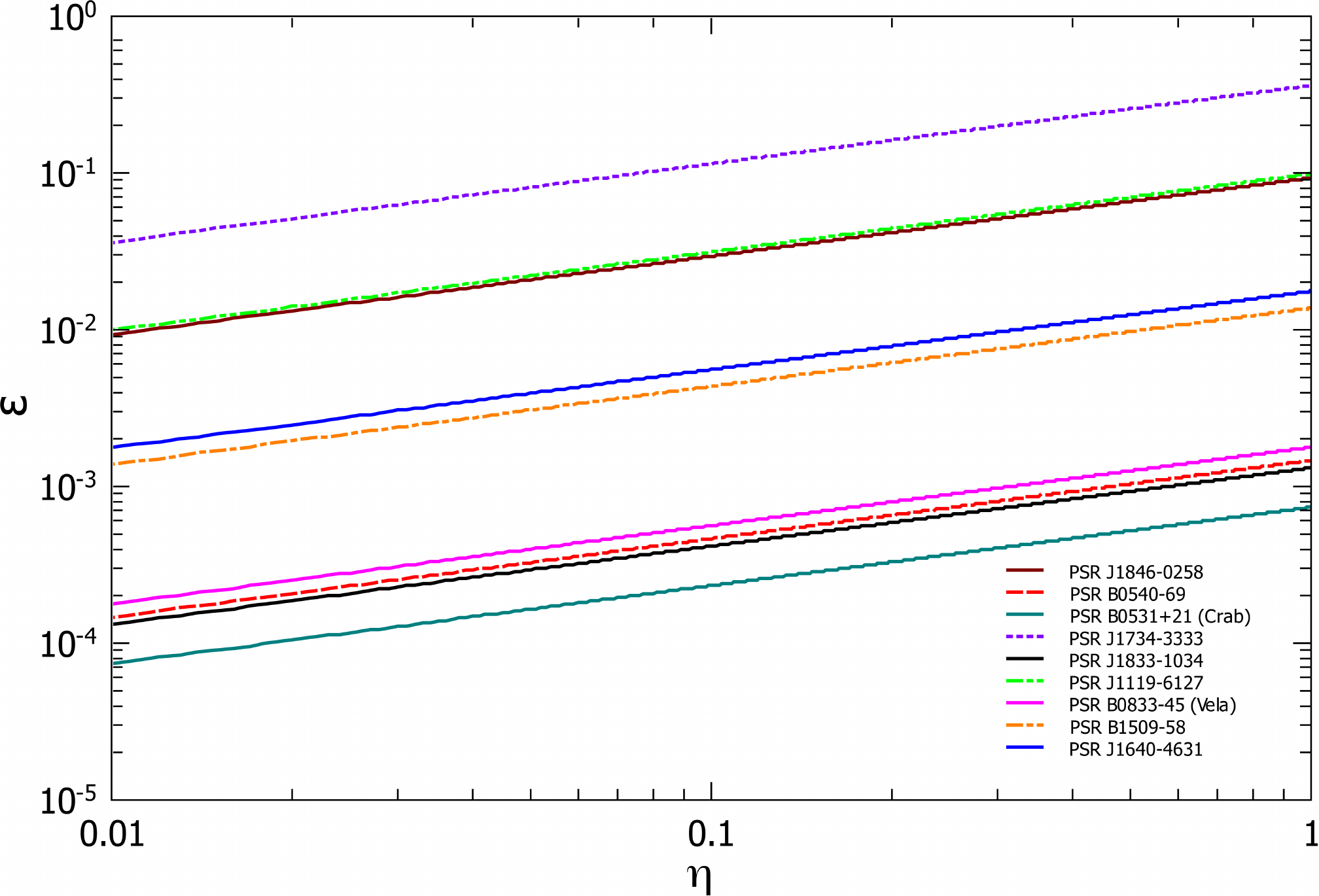} 
\caption{$\epsilon$ as a function of $\eta$.}\label{fig2}
\end{figure}

\begin{figure}
\includegraphics[width=\linewidth,clip]{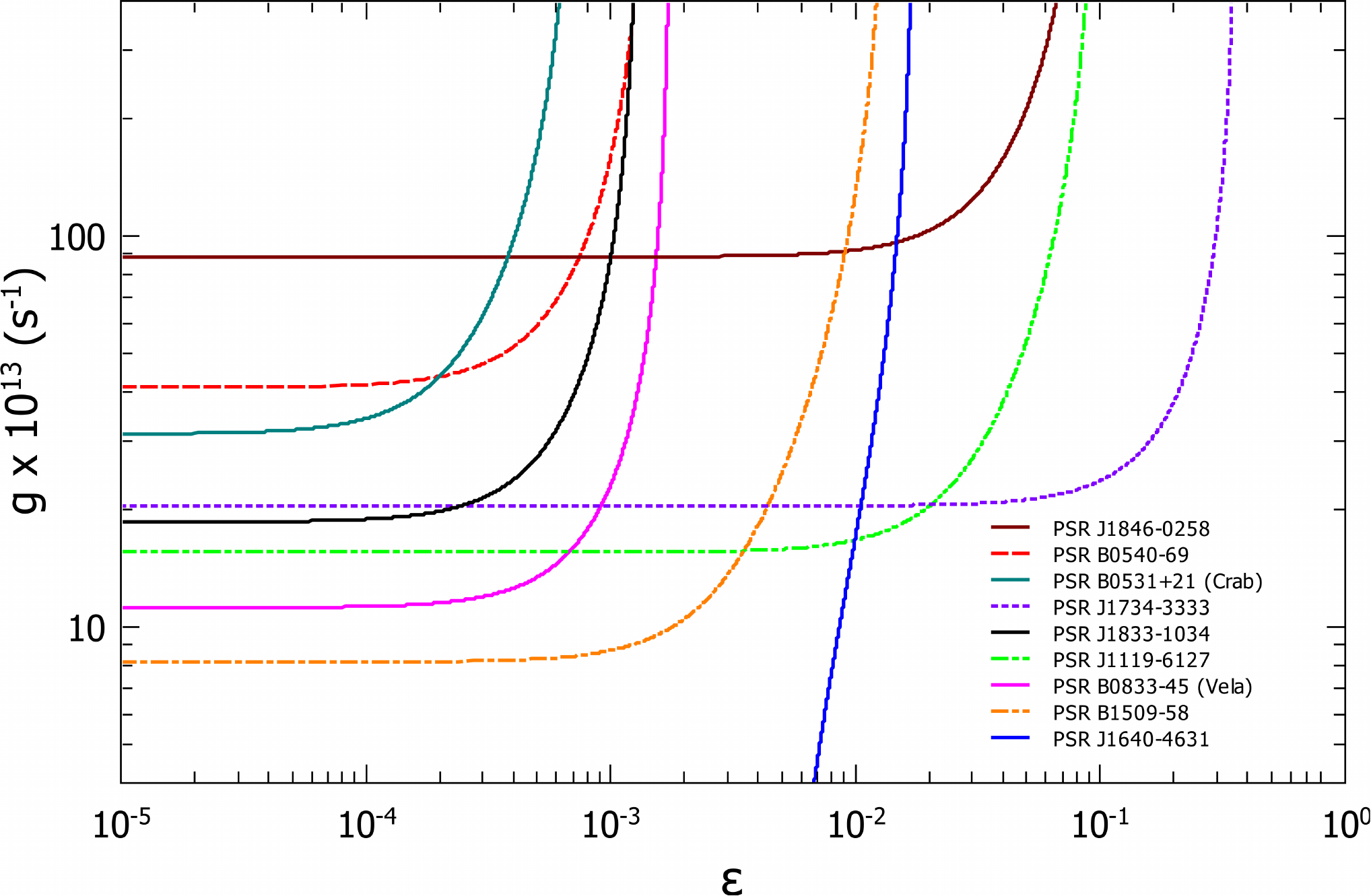} 
\caption{$g$ as a function of $\epsilon$.}\label{fig3}
\end{figure}

From equation \ref{eta0} we can write
\begin{equation}
\dot{\bar{f}}_{\rm rot} = \eta \dot{f}_{\rm rot}, 
\end{equation}
where $\dot{\bar{f}}_{\rm rot}$ can be interpreted as the part of $\dot{f}_{\rm rot}$ related to the GW emission brake. Thus, the amplitude of the GWs is now given by
\begin{equation}
\bar{h}^2 = \frac{5}{2}\frac{G}{c^3}\frac{I}{r^2}\frac{\mid\dot{\bar{f}}_{rot}\mid}{f_{rot}} =  \frac{5}{2}\frac{G}{c^3}\frac{I}{r^2}\frac{\mid\dot{f}_{\rm rot}\mid}{f_{\rm rot}} \, \eta . \label{heta}
\end{equation}
On the other hand, recall that the amplitude of GWs can also be written as follows
\begin{equation}
h = \frac{16\pi^2G}{c^4} \frac{I\epsilon f_{\rm rot}^2}{r},
\end{equation}
\citep[see, e.g.,][]{1983bhwd.book.....S}, which with the use of equation \ref{heta} yields an equation for $\epsilon$ in terms of $P$, $\dot P$ (observable quantities), $\eta$ and $I$, namely
\begin{equation}
\epsilon = \sqrt{\frac{5}{512\pi^4} \frac{c^5}{G}\frac{\dot{P}P^3}{I}\eta}. \label{epet}
\end{equation}
Assuming that $I \approx 10^{38} \, \rm{kg \, m^2}$ (fiducial value) and substituting the values of $P$, $\dot P$ from the pulsars listed in Table \ref{ta1}, we obtain $\epsilon$ as a function of $\eta$ for these pulsars. Figure \ref{fig2} shows $\epsilon$ as a function of $\eta$.
Notice that even for an efficiency of GW generation ($\eta$) of $1\%$, $\epsilon$ is relatively large. This could well be an indication that $\eta \ll 0.01$ for some pulsars of Table \ref{ta1}, if it is required that $\epsilon < 10^{-3}$.  Equation \ref{epet} also allows one to write the equation for $g$ now as function of $\epsilon$. In figure \ref{fig3} we just show a graph thereof.

At this point it is interesting to see what kind of information we can obtain from \cite{2014ApJ...785..119A}, since these authors studied GWs from known pulsars. In such a paper the authors searched for GWs from 195 pulsars from the LIGO and Virgo S3/S4, S5, S6, VSR2, and VSR4 runs. Although they did not
find any evidence for GWs, it was possible to provide upper limits to the GW amplitudes. Moreover, using different statistical methods, they could also find upper limits for $\epsilon$ and $\eta$. Also, they pointed out seven pulsars of high interest. Notice that only three (Vela, Crab and PSR J1833-1034) of the nine pulsars here studied (see Table \ref{ta1}) were studied in Ref. \cite{2014ApJ...785..119A}. For Crab and Vela, their results are consistent with $\eta \sim 0.01$ and $\epsilon \sim 10^{-4}$, just like ours. For PSR J1833-1034, however, their results indicated an inconsistent value of $\eta \simeq 15-20$, since this efficiency violates the energy balance. On the other hand, our results indicate that for $\epsilon \ll 10^{-4}$, $\eta \ll 1 \% $ for this very same pulsar, which is a more acceptable value. It is worth mentioning, that the above comparison is limited to the consistency of the results, since these authors are considering statistical tools to find upper limits for a series of parameters, we instead perform an analytical approach. 

\begin{figure}
\includegraphics[width=\linewidth,clip]{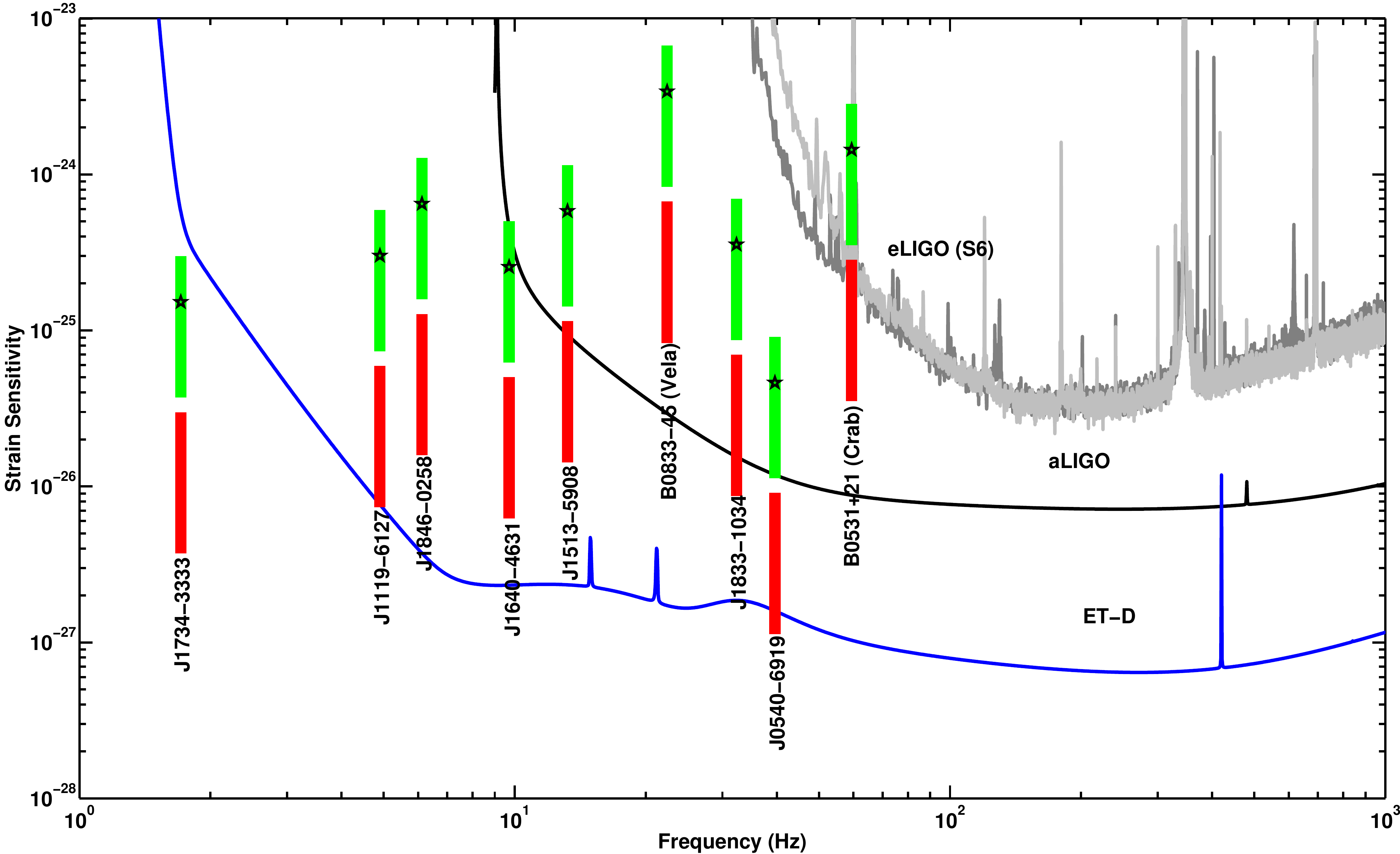} 
\caption{Amplitude of the GWs for the pulsars listed in Table \ref{ta1} as a function of the GW frequency, $f_{\rm GW} = 2f_{\rm rot}$, for $\eta=0.01$ (red bars) and $\eta=1$ (green bars)
for ${7\times 10^{36}\lesssim I \lesssim 3\times 10^{38}\,{\rm kg.m^2}}$ 
(see the text for further details). Also plotted the strain sensitivities for eLIGO S6, aLIGO and ET-D for one-year integration time. The stars represent such amplitudes when they are calculated for $\eta=1$ and $I_{\rm fiducial}=10^{38}{\rm kg.m^2}$.}\label{fig4}
\end{figure}

Now, we calculate the amplitudes of the GWs generated by the pulsars considered in this present paper. Figure \ref{fig4} shows the amplitudes of the GWs generated by the pulsars listed in Table \ref{ta1} as a function of the GW frequency, $f_{\rm GW} = 2f_{\rm rot}$.
In doing this we use a range of neutron star masses in equation~\ref{heta}, using selected up-to-date nuclear equations of state (EOS). In particular, for the sake of simplicity, we adopt the GM1 parametrization~\citep{1991PhRvL..67.2414G}. Figure \ref{fig4} shows a range of possible amplitudes (for two distinct efficiencies, $\eta=0.01$ and $\eta=1$, represented by red and green bars respectively) calculated from the realistic moment of inertia ($7.03\times 10^{36}\leq I \leq 3.28\times 10^{38}$) $\rm kg.m^2$. 
Also plotted the strain sensitivities for eLIGO S6, aLIGO and ET-D for one-year integration
time~\citep{2015CQGra..32g4001L, 2015CQGra..32k5012A,2011CQGra..28i4013H}.
\section{Discussion and final remarks}\label{sec4}
In this paper we study the putative emission of GWs generated in particular by pulsars with measured braking indices. We model the braking indices of these pulsars taking into account in the spindown
the magnetic dipole and GWs brakes, besides the surface magnetic dipole and the angle between the magnetic and rotation axes dependent on time. We show that the appropriate combination of these very quantities can account for the braking indices observed for pulsars listed in Table~\ref{ta1}.

Notice that even for $\eta = 0.01$ our model predicts that some pulsars would have large values of $\epsilon$. This can be an indication that $\eta \ll 0.01$ for these pulsars, implying in smaller ellipticities, i.e., $\epsilon < 10^{-3}$.

Moreover, we study the detectability of these pulsars by aLIGO and ET. Besides considering the efficiency effect of GWs generation, we take into account the role of the moment of inertia. To do so we take into account different values for $I$ that come from models of neutron stars for a given and acceptable equation of state.

We show that aLIGO can well detect at least some of the pulsars considered here in particular Vela within one year of observation. 

Last, but not least, even if one studies scenarios in which alternative models to explain the pulsars'  braking index are considered, it is quite important to include the putative contribution of GWs. Since it is quite reasonable to expect that pulsars have deformations, which implies in a finite value for the ellipticity, which is one of the main quantities in the calculation of the GW amplitudes. On the other hand, from the point of view of the searches for continuous GWs, there is a quite recently published paper   \cite{2016PhRvD..93f4011M} with interesting strategies in which not only GWs are considered as emission mechanisms.

\begin{acknowledgements}
J.C.N.A thanks FAPESP (2013/26258-4) and CNPq (308983/2013-0) for partial support.
J.G.C. acknowledges the support of FAPESP (2013/15088-0 and 2013/26258-4).  C.A.C. acknowledges PNPD-CAPES for financial support. We would also like to thank the referee for his (her) useful suggestions and criticisms. 
\end{acknowledgements}

%\bibliography{ref}   % name your BibTeX data base

\end{document}